\documentclass[fleqn,11pt]{article}
\usepackage[cp1251]{inputenc}
\usepackage{amsmath,amssymb}

\usepackage{amsfonts,amssymb,cite}
\usepackage{graphicx}

\def\noi{\noindent}

\newcommand{\Title}[1]{\noi {{\Large\bf #1}}\\[1ex]}

\newcommand{\Author}[2]{\noi{\bf #1}\\[2ex]\noi{\normalsize\it #2}\\}

\newcommand{\Abstract}[1]{\vskip 2mm \begin{center}
        \parbox{16.4cm}{\small\noi #1} \end{center}\medskip}
\newcommand{\foom}[1]{\protect\footnotemark[#1]}

\def\nqq{\hspace*{-2em}}

\def\Jl#1#2{#1 {\bf #2},\ }

\def\ApJ#1 {\Jl{Astroph. J.}{#1}}
\def\CQG#1 {\Jl{Class. Quantum Grav.}{#1}}
\def\DAN#1 {\Jl{Dokl. AN SSSR}{#1}}
\def\GC#1 {\Jl{Grav. Cosmol.}{#1}}
\def\GRG#1 {\Jl{Gen. Rel. Grav.}{#1}}
\def\JETF#1 {\Jl{Zh. Eksp. Teor. Fiz.}{#1}}
\def\JETP#1 {\Jl{Sov. Phys. JETP}{#1}}
\def\JHEP#1 {\Jl{JHEP}{#1}}
\def\JMP#1 {\Jl{J. Math. Phys.}{#1}}
\def\NPB#1 {\Jl{Nucl. Phys. B}{#1}}
\def\NP#1 {\Jl{Nucl. Phys.}{#1}}
\def\PLA#1 {\Jl{Phys. Lett. A}{#1}}
\def\PLB#1 {\Jl{Phys. Lett. B}{#1}}
\def\PRD#1 {\Jl{Phys. Rev. D}{#1}}
\def\PRL#1 {\Jl{Phys. Rev. Lett.}{#1}}

\def\lal{&&\nqq {}}

\def\beq{\begin{equation}}
\def\eeq{\end{equation}}
\def\bear{\begin{eqnarray}}
\def\bearr{\begin{eqnarray} \lal}
\def\ear{\end{eqnarray}}
\def\earn{\nonumber \end{eqnarray}}

\newcommand{\Fig}[3]{%
\begin{center}
\parbox{8cm}{%
\refstepcounter{figure}\includegraphics[width=8cm,height=#2cm]{#1} \noindent Figure \thefigure:\quad
#3}\end{center}}

\begin{document}
\thispagestyle{empty}
\twocolumn[

\Title{The Stability of the Cosmological System of Degenerate Scalar Charged Fermions and Higgs Scalar Fields. II. The Evolution of Short-Wave Perturbations.\foom 1}

\Author{Yu.G. Ignat'ev}
    {Institute of Physics, Kazan Federal University, Kremlyovskaya str., 18, Kazan, 420008, Russia}


\Abstract
 {A mathematical model of the evolution of plane perturbations in the cosmological statistical system of completely degenerate scalar-charged fermions with Higgs scalar interaction for short-wave perturbations is formulated. These disturbance modes have been found and investigated. It is shown that in such a model an additional oscillation mode arises, which is associated directly with degenerate fermions, in which monotonically decreasing and increasing perturbations of the metric exist. In this case, there is such a time instant in the system that the potential of the scalar field, together with the energy density, tends to infinity. In this regard, an assumption is made about a possible mechanism for the formation of dark matter condensates.
}
\bigskip

] 
\section{Equations for Plane Perturbations}
In the previous part of the work  \cite{YuI_20_1part} a mathematical model of plane perturbations in the cosmological scalar-charged cosmological system of degenerate fermions interacting through the Higgs scalar field was formulated.

\subsection{The Background Equations}
The model consists, first, of nonlinear background equations describing a homogeneous isotropic cosmological system of degenerate scalar charged fermions \cite{YuI_20} on the background of the spatially flat Friedmann metric:
\begin{equation}\label{ds_0}
ds_0^2=a^2(\eta)(d\eta^2-dx^2-dy^2-dz^2).
\end{equation}

\emph{1. The Background Scalar Field Equation} $\Phi$ in the Friedmann metric:
\begin{equation}\label{Eq_S_eta}
\mathrm{e}\biggl(\ddot{\Phi}+2\frac{\dot{a}}{a}\dot{\Phi}\biggr)+a^2(m^2\Phi-\alpha\Phi^3)= -8\pi a^2\sigma(\eta),
\end{equation}
where
\begin{equation}\label{sigma_0}
\sigma(\eta)=\frac{q^4\Phi^3(\eta)}{2\pi^2}F_1(\psi(\eta))
\end{equation}
-- is the scalar charge's density, $e=\pm 1$, ``$+$'' corresponds to a scalar field, ``$-$'' -- to a phantom one; $m$ -- is the Higgs bosons' mass, $\alpha$ - is a self-action constant, $q$ -- is a scalar charge of fermions.

\emph{2. The Einstein equations for the Friedmann metric}
\begin{eqnarray}\label{Surf_Einst}
3H^2-\tilde{\Lambda}-\frac{q^4\Phi^4}{\pi}F_2(\psi)-\nonumber\\
\frac{e\dot{\Phi}^2}{2a^2}-\frac{m^2\Phi^2}{2}+\frac{\alpha\Phi^4}{4}=0;
\end{eqnarray}
\begin{equation}\label{11-44}
\dot{H}+\frac{e\dot{\Phi}^2}{2a}+\frac{4 a}{3\pi}m^4_*\psi^3\sqrt{1+\psi^2}=0,
\end{equation}
where $H(\eta)$ -- is the Hubble constant,
\begin{eqnarray}\label{H}
H= \frac{\dot{a}}{a^2},\\
\label{tilde{Lambda}}
\tilde{\Lambda}=\Lambda-\frac{m^4}{4\alpha},
\end{eqnarray}
$\Lambda$ -- is a cosmological constant.

Further, for the dimensionless function $\psi$, which is equal to the ratio of the Fermi energy to the effective rest mass of fermions, the following expression is obtained in \cite{YuI_20}
\begin{equation}\label{psi(eta)}
\psi(\eta)=\frac{\beta}{a(\eta)\Phi(\eta)},\quad \biggl(\beta=\frac{\varrho}{q},\; a(0)\equiv=1\biggl),
\end{equation}
where $\varrho$ is a Fermi energy at $\eta=0$. Finally, the functions $F_1(x),F_2(x)$ are equal to:
\begin{eqnarray}
\!\! F_1(\psi)=\psi\sqrt{1+\psi^2}-\ln(\psi+\sqrt{1+\psi^2});\nonumber\\
\!\! F_2(\psi)=\psi\sqrt{1+\psi^2}(1+2\psi^2)-\ln(\psi+\sqrt{1+\psi^2}).\nonumber
\end{eqnarray}
This background model was investigated qualitatively and numerically in  \cite{YuI_20}.

\subsection{First-order Equations of Perturbation Theory for Plane Perturbations}
Let us define the perturbed values of the dynamic quantities for small plane perturbations of the form $S(\eta)\mathrm{e}^{inz}$, propagating in the direction of the $Oz$ axis, --
\begin{eqnarray}\label{dF-drho-du}
\Phi(z,\eta)\rightarrow\Phi(\eta)+\varphi(\eta)\mathrm{e}^{inz};&\nonumber\\
\varrho(z,t)\rightarrow\varrho(\eta)+\delta(\eta)\mathrm{e}^{inz};&\\
\sigma(z,\eta)\rightarrow \sigma(\eta)+s(\eta)\mathrm{e}^{inz};&\nonumber\\
u^i=\frac{1}{a}\delta^i_4+\delta^i_3 v(\eta)\mathrm{e}^{inz},&\nonumber
\end{eqnarray}
where $\varphi(\eta),\delta(\eta),s(\eta),v(\eta)$ -- are the functions of the first order of smallness in comparison with their unperturbed values where $\delta(\eta)$ -- is a perturbation of the Fermi momentum $\varrho(\eta)$. For such perturbations, in \cite{YuI_20_1part}  the following system of ordinary linear differential equations was obtained:
\begin{eqnarray}\label{Eq_dphi}
e\ddot{\varphi}+2\frac{\dot{a}}{a}e\dot{\varphi}+\bigl[e n^2+a^2(m^2-3\alpha\Phi^2)\bigr]\varphi\nonumber\\
+\frac{e}{2}\dot{\Phi}\dot{\mu}=-8\pi a^2s.\\
\label{34}
v=\frac{in}{8\pi a^3(\varepsilon+p)_p}\biggl(e\varphi\dot{\Phi}+\frac{1}{3}(\dot{\lambda}+\dot{\mu})\biggr);\\
\label{44}
8\pi a^2\delta\varepsilon_p=\frac{\dot{a}}{a}\dot{\mu}-e\dot{\Phi}\dot{\varphi}+\frac{n^2}{3}(\lambda+\mu)\nonumber\\
-a^2(m^2-\alpha\Phi^2)\Phi\varphi;\\
\label{11-33}
\ddot{\lambda}+2\frac{\dot{a}}{a}\dot{\lambda}-\frac{1}{3}n^2(\lambda+\mu)=0;\\
\label{11+22+33}
\ddot{\mu}+2\frac{\dot{a}}{a}\dot{\mu}+\frac{1}{3}n^2(\lambda+\mu)+3e\dot{\varphi}\dot{\Phi}+\nonumber\\
-3 a^2(\Phi\varphi(m^2-\alpha\Phi^2)-8\pi\delta p_p)=0.
\end{eqnarray}
\subsection{The Perturbations of the Fermi Component}
Let us find, introducing the new function
\begin{equation}\label{gamma}
\gamma(\eta)=\delta(\eta)-\varphi(\eta),
\end{equation}
the relation for the perturbation of the relative energy of fermions:
\begin{eqnarray}\label{dpsi}
\delta\psi=\psi(\eta)\gamma(\eta)\mathrm{e}^{inz}.
\end{eqnarray}
The perturbations of macroscopic scalars of fermions are completely determined by two scalar functions $\varphi(\eta)$ and $\gamma(\eta)$ \cite{YuI_20_1part}:
\begin{equation}\label{s0}
s =\frac{q^4\Phi^3(\eta)}{2\pi^2}\biggl(3\varphi F_1(\psi) +\gamma\frac{\psi^3}{\sqrt{1+\psi^2}}\biggr);\end{equation}
\begin{equation}\label{de0}\delta\varepsilon_p=\frac{q^4\Phi^4\psi}{2\pi^2}\bigl(\varphi F_2(\psi)+2\gamma\psi^2\sqrt{1+\psi^2}\bigl);\end{equation}
\begin{equation}\label{dp0}\delta p_p=\frac{q^4\Phi^4}{6\pi^2}\biggl[\varphi F_2(\psi)+\gamma\frac{\psi^3(2+\psi^2)}{\sqrt{1+\psi^2}}\biggl] ;\end{equation}
\begin{equation}\label{d(e-3p0)}\delta(\varepsilon-3p)_p=\frac{q^4\Phi^3}{2\pi^2}\biggl(4\varphi F_1(\psi)+\gamma\frac{\Phi\psi^3}{\sqrt{1+\psi^2}}
\biggr).\end{equation}

As a result, using the relation \eqref{de0} in the equation \eqref{44} and assuming
\begin{equation}\label{Phi_not_0}
\Phi\not\equiv0,
\end{equation}
we find an expression for the perturbation of the relative Fermi energy \eqref{dpsi} $\gamma(\eta)$ through the perturbations of the fields $\varphi,\lambda,\mu$ and their first derivatives:
\begin{eqnarray}\label{gamma=}
\!\!\!\!\gamma=-\frac{1}{2\psi^2\sqrt{1+\psi^2}}\biggl[\varphi\biggr(F_2(\psi)+\frac{\pi(m^2-\alpha\Phi^2)}{4q^4\Phi^3\psi}\biggl)&\nonumber\\
\!\!\!\!\ -\frac{\pi n^2(\lambda+\mu)}{12 q^4\Phi^4\psi a^2}-\frac{\pi}{4}\frac{\dot{\mu}\dot{a}}{q^4\Phi^4\psi a^3}+\frac{e\pi}{4}\frac{\dot{\Phi}\dot{\varphi}}{q^4\Phi^4\psi a^2}\biggr].&\nonumber\\
&
\end{eqnarray}

Substituting \eqref{gamma=} into the relations \eqref{s0} and \eqref{dp0}, we get:
\begin{eqnarray}\label{s}
s =\!\biggl[\frac{q^4\Phi^3}{2\pi^2}\biggl(3F_1(\psi)-\frac{\psi F_2(\psi)}{2(1+\psi^2)}\biggr)\nonumber\\
-\frac{m^2-\alpha\Phi^2}{16\pi(1+\psi^2) } \biggr]\varphi +\frac{n^2(\lambda+\mu)}{48\pi(1+\psi^2)\Phi a^2}-\nonumber\\
\frac{e\dot{\Phi}\dot{\varphi}}{16\pi\Phi(1+\psi^2)a^2}+\frac{\dot{\mu}\dot{a}}{16\pi\Phi(1+\psi^2)a^3};
\end{eqnarray}
\begin{eqnarray}\label{dp}
\delta p_p=\frac{1}{6\pi^2}\biggl[F_2(\psi)q^4\Phi^4\biggl(1-\frac{1}{2}\psi\chi(\psi)\biggr) \nonumber\\
-\frac{\pi\Phi}{8}(m^2-\alpha\Phi^2)\chi(\psi)\biggr]\varphi +\frac{\chi(\psi)}{144\pi}\frac{n^2(\lambda+\mu)}{a^2}\nonumber\\
-\frac{e\chi(\psi)}{48\pi}\frac{\dot{\varphi}\dot{\Phi}}{a^2}+\frac{e\chi(\psi)}{48\pi}\frac{\dot{a}\dot{\mu}}{a^3},
\end{eqnarray}
where the following weakly varying function is introduced
\begin{equation}\label{chi}
\chi(\psi)=\frac{2+\psi^2}{1+\psi^2}, \quad 1\leqslant\chi(\psi)\leqslant 2.
\end{equation}
\subsection{Complete closed system of equations for perturbations}
Adding equations \eqref{11-33} and \eqref{11+22+33} and substituting the expression for $\delta p_p$ into the obtained equation from \eqref{dp} and introducing a new perturbation amplitude of the metric %
\begin{equation}\label{nu}
\nu=(\lambda+\mu),
\end{equation}
we get an equation with respect to this quantity:
\begin{eqnarray}\label{Eq_nu}
\ddot{\nu}+2\frac{\dot{a}}{a}\dot{\nu}+\frac{n^2}{6}\chi(\psi)\nu+\frac{a^2}{4\pi}\biggl[q^4\Phi^4F_2(\psi)\biggl(1-\nonumber\\
\bar{}\frac{1}{2}\psi\chi(\psi)\biggr)-\biggl(12+\frac{\chi(\psi)}{8}\biggr)\pi\Phi(m^2-\alpha\Phi^2)\biggr]\varphi\nonumber\\
+e\biggl(3-\frac{\chi(\psi)}{2}\biggr)\dot{\varphi}\dot{\Phi}+\frac{\chi(\psi)}{2}\frac{\dot{a}}{a}(\dot{\nu}-\dot{\lambda})=0.
\end{eqnarray}
Substituting the expression for $s$ from \eqref{s} into the equation \eqref{Eq_dphi}, we obtain the equation for the perturbation of the scalar potential:
\begin{eqnarray}\label{Eq_varphi}
e\ddot{\varphi}+2e\frac{\dot{a}}{a}\dot{\varphi} +\biggl[en^2+a^2(m^2-3\alpha\Phi^2)+\nonumber\\
\frac{4a^2q^4\Phi^3}{\pi}\biggl(3F_1(\psi)-\frac{F_2(\psi)}{2(1+\psi^2)}\biggr)-\nonumber\\
\frac{a^2(m^2-\alpha\Phi^2)}{2(1+\psi^2)}\biggr]\varphi+\frac{n^2\nu}{6\Phi(1+\psi^2)}\nonumber\\
 -\frac{e\dot{\Phi}\dot{\varphi}}{2\Phi(1+\psi^2)}+\frac{\dot{a}(\dot{\nu}-\dot{\lambda})}{a^2\Phi(1+\psi^2)}=0.
\end{eqnarray}

Finally, using the relation \eqref{nu}, we rewrite the equation \eqref{11-33} in new variables
\begin{equation}\label{Eq_lambda}
\ddot{\lambda}+2\frac{\dot{a}}{a}\dot{\lambda}-\frac{1}{3}n^2\nu=0.
\end{equation}

Thus, we have a complete system of three linear ordinary differential equations  \eqref{Eq_varphi},  \eqref{Eq_lambda} and \eqref{Eq_nu} with respect to three functions $\varphi, \lambda$ and $\nu$.

\section{The WKB Approximation}
\subsection{The Complete System of Equations for Perturbations in the Hard WKB Approximation}
Let us investigate the formulated mathematical model of the cosmological evolution of perturbations  \eqref{Eq_dphi} -- \eqref{gamma=} in the short-wave sector of perturbations
\begin{eqnarray}\label{WKB}
n\gg \biggl\{\frac{\dot{a}}{a},\frac{\dot{\Phi}}{\Phi}\biggr\};\qquad \dot{\varphi}\gg \frac{\dot{a}}{a}\varphi
\end{eqnarray}
In addition, in this article we will restrict ourselves to the study of the \emph{hard WKB - approximation}, assuming
\begin{equation}\label{extrem_WKB} n\gg \mathrm{sup}\{am,a|\Phi|,aq^2|\Phi|\}.
\end{equation}
Let us notice that in this hard WKB approximation we lose the terms in the equations for the perturbations that are responsible for the specifics of the Higgs interaction. This specifics will affect the evolution of perturbations only by means of the time dependence of the background quantities $a(\eta)$ and $\Phi(\eta)$. We intend to return to considering corrections to the evolutionary equations for perturbations determined by the Higgs interaction in the next article.

In accordance with the WKB method, we represent solutions of the equations $f(\eta)$ in the form
\begin{equation}\label{Eiconal}
f=\tilde{f}(\eta) \cdot \mathrm{e}^{i\int u(\eta)d\eta}; \quad (|u|\gg 1),
\end{equation}
where $\tilde{f}(\eta)$ and $u(\eta)$ -- are functions of the \emph{amplitude} and \emph{eikonal} of the perturbation slightly changing together with the scale factor, such that:
\begin{equation}\label{WKB1}
\frac{\dot{a}}{a}\sim \frac{1}{\ell};\; \dot{\tilde{f}}\sim \frac{\tilde{f}}{\ell}; \; \dot{u}\sim \frac{u}{\ell}.
\end{equation}
Thus, we will assume the below quantities are large
\begin{equation}\label{nl>>1}
n\ell\gg 1;\quad u\ell \gg 1
\end{equation}
and take into account the hard WKB approximation \eqref{extrem_WKB}. Keeping the terms up to the first order of the WKB expansion, we bring the studied system of equations  \eqref{Eq_varphi}, \eqref{Eq_lambda} and \eqref{Eq_nu} to the form (let us note that the equation \eqref{Eq_varphi} retains its form):
\begin{eqnarray}\label{Eq_varphi_WKB}
e\ddot{\varphi}+2\frac{\dot{a}}{a}\dot{\varphi} +\frac{n^2\nu}{6\Phi(1+\psi^2)}-\nonumber\\
\frac{e\dot{\Phi}\dot{\varphi}}{2\Phi(1+\psi^2)}+\frac{\dot{a}(\dot{\nu}-\dot{\lambda})}{a^2\Phi(1+\psi^2)}=0;
\end{eqnarray}
\begin{equation}\label{Eq_lambda_WKB}
\ddot{\lambda}+2\frac{\dot{a}}{a}\dot{\lambda}-\frac{1}{3}n^2\nu=0.
\end{equation}
\begin{eqnarray}\label{Eq_nu_WKB}
\ddot{\nu}+\frac{n^2}{6}\chi(\psi)\nu+2\frac{\dot{a}}{a}\dot{\nu}+e\biggl(3-\frac{\chi(\psi)}{2}\biggr)\dot{\varphi}\dot{\Phi}\nonumber\\
+\frac{\chi(\psi)}{2}\frac{\dot{a}}{a}(\dot{\nu}-\dot{\lambda})=0.
\end{eqnarray}
\subsection{Expansion of Equations in WKB Orders}
 Let us substitute perturbations in the form \eqref{Eiconal} into the equations \eqref{Eq_varphi_WKB} -- \eqref{Eq_nu_WKB} and then expand them in orders of large value $ul$ \eqref{WKB}, limiting ourselves to the first order of WKB approximation. Thus, we obtain the WKB equations of the zero and first order.

\textbf{Zero Order:}
\begin{eqnarray}
\label{varphi_0}
e(-u^2+n^2)\tilde{\varphi}+\frac{n^2}{6\Phi(1+\psi^2)}\tilde{\nu}=0;\\
\label{l,nu_0}
-u^2\tilde{\lambda}-\frac{n^2}{3}\tilde{\nu}=0;\\
\label{nu_0}
-u^2\tilde{\nu}+\frac{n^2}{6}\chi(\psi)\tilde{\nu}=0.
\end{eqnarray}

\textbf{First Order:}
\begin{eqnarray}
\label{dot(varphi)}
2u\dot{\tilde{\varphi}}+2\frac{\dot{a}}{a}u\tilde{\varphi}+\dot{u}\tilde{\varphi}-\nonumber\\
\frac{u\dot{\Phi}\tilde{\varphi}}{2\Phi(1+\psi^2)}+
\frac{\dot{a}u(\tilde{\nu}-\tilde{\lambda})}{a^2\Phi(1+\psi^2)}=0;\\
\label{dot(lambda)}
2u\dot{\tilde{\lambda}}+\dot{u}\tilde{\lambda}+2\frac{\dot{a}}{a}u\tilde{\lambda}=0;\\
\label{dot(nu)}
2u\dot{\tilde{\nu}}+\dot{u}\tilde{\nu}+2\frac{\dot{a}}{a}u\tilde{\nu}+\nonumber\\
eu\tilde{\varphi}\biggl(3-\frac{1}{2}\chi(\psi)\biggr)\dot{\Phi}+\frac{\dot{a}}{2a}u\chi(\psi)(\tilde{\nu}-\tilde{\lambda})=0.
\end{eqnarray}

Let us note that the equation  \eqref{dot(lambda)} has its integral:
\begin{equation}\label{lambda_1}
\tilde{\lambda}=\frac{\mathrm{Const}}{a\sqrt{u}}.
\end{equation}

\subsection{The WKB - solution}
 The necessary and sufficient condition for the non-trivial consistency of the system of linear homogeneous algebraic equations \eqref{varphi_0} -- \eqref{nu_0} with respect to the functions $\{\varphi,\lambda,\nu\}$ is the determinant of this system being equal to zero
\[\Delta=eu^2(-u^2+n^2)\biggl(u^2-\chi(\psi)\frac{n^2}{6}\biggr).\]
Thus, we have the following solutions for the eikonal function:
\begin{eqnarray}
\label{u(eta)}
\!\! u_{1,2}=0; \; u_{3,4}=\pm n;\; u_{5,6}=\pm \frac{\pi n}{\sqrt{6}}\chi^{1/2}(\psi)=0.
\end{eqnarray}

Substitution of the zero solution $u_{1,2}$ into the original system \eqref{varphi_0} -- \eqref{nu_0} gives us the trivial solution $\tilde{\varphi}=\tilde{\lambda}=\tilde{\nu}=0$. Substitution of the solution $u_{3,4}$ into the original system gives $\tilde{\lambda}=\tilde{\nu}=0$, retaining the function's $\tilde{\varphi}$ arbitrariness. This function is determined from the first order equation of the WKB approximation \eqref{dot(varphi)}, which gives:
\begin{equation}\label{varphi_3,4}
\varphi_{1,2}=\frac{C_\pm}{a(\eta)}\mathrm{e}^{\pm in\eta}.
\end{equation}
Let us note that exactly this solution for the classical field ($e=+1$)  was obtained in \cite{YuI_20} or the vacuum Higgs model.

Finally, substitution of the $u_{5,6}$ solution into the original system gives a new relationship between the required amplitudes:
\begin{eqnarray}\label{new_amp_nu}
\tilde{\nu}=-\frac{\chi(\psi)}{2}\tilde{\lambda};\\
\label{new_amp_varphi}
\tilde{\varphi}=\frac{e\chi(\psi)}{12\Phi(1+\psi^2)}\biggl(1-\frac{\chi(\psi)}{6}\biggr)^{-1}\tilde{\lambda}.
\end{eqnarray}

Substituting the function  $\tilde{\lambda}$ from \eqref{lambda_1} and then the solutions $u_{5,6}$ for the eikonal function into the relations \eqref{new_amp_nu} -- \eqref{new_amp_varphi}we obtain the new solution:
\begin{eqnarray}
\label{varphi_sol}
\varphi=\frac{C_\pm e\chi(\psi)}{12 a\Phi(1+\psi^2)}\biggl(1-\frac{\chi(\psi)}{6}\biggr)^{-1} U_\pm(\eta); \\
\label{lambda_sol}
\lambda=\frac{C_\pm}{a}U_\pm(\eta) ;\\
\label{nu_sol}
\nu=-\frac{C_\pm\chi(\psi)}{2a} U_\pm(\eta),
\end{eqnarray}
where
\begin{equation}\label{Upm}
U_\pm(\eta)=\exp\biggl(\pm i\frac{\pi n}{\sqrt{6}}\int \sqrt{\chi(\psi)}d\eta\biggr).
\end{equation}

\section{The Solution Analysis}
Let us remind that, first of all,  the `` hard '' WKB approximation \eqref{extrem_WKB} onsidered in this article does not explicitly take into account the specifics of the Higgs interaction in the formulas for the evolution of perturbations \eqref{varphi_sol} -- \eqref{Upm}. We intend to take this specifics into account in the next article. In the approximation \eqref{extrem_WKB} the specificity of the Hinggs interaction affects only the background functions  $\Phi(\eta)$ and $a(\eta)$, which determine the time dependence of the solution\eqref{varphi_sol} -- \eqref{Upm}. Let us remind that the corresponding solution can be obtained only in numerical form (see \cite{YuI_20}).

Postponing the solution of these problems for the future, we nevertheless can draw some important conclusions regarding the evolution of short-wavelength perturbations in a homogeneous isotropic cosmological system of degenerate scalar charged fermions. Leaving aside the standard perturbation mode  \eqref{varphi_3,4}, which is not accompanied by perturbations of the metric, let us first of all, note that  taking into account the weak time dependence of the function $\chi(\psi)$ \eqref{chi} and the connection of the time variable $\eta$ with physical time
\[dt= a(\eta)d\eta\]
we obtain the following relation for the oscillation phase $\phi$ from \eqref{Upm}
\[\phi=nz\pm \frac{\pi n}{\sqrt{6}}\int\sqrt{\chi(\psi)}d\eta,\]
differentiating which we find for the phase velocity of oscillations:
\[v_{ph}=\frac{\pi\sqrt{\chi(\psi)}}{\sqrt{6}a}.\]
This mode of oscillations, in contrast with the vacuum one \eqref{varphi_3,4}, is connected with fermion oscillations.

Then, from the formulas  \eqref{lambda_sol} -- \eqref{nu_sol} taking into account \eqref{nu} we see that the amplitudes of the perturbation of the metric $\lambda(\eta)$ and $\mu(\eta)$ in this mode of perturbation monotonously decrease with time as $a^{-1}$. The behavior of the perturbation of the potential of the scalar field $\varphi(\eta)$ according to \eqref{varphi_sol}, is mainly determined by the factor $a^{-1}(\eta)\Phi^{-1}(\eta)$, that is, by the background fields that can only be found by means of numerical integration of the background equations \eqref{Eq_S_eta} -- \eqref{11-44}. From \eqref{varphi_sol} we see that the behavior of scalar field perturbations for the classical and phantom fields formally differ only by the factor $e=\pm1$. However, the behavior of background solutions for classical and phantom fields is fundamentally different (see \cite{YuI_20}). In particular, the behavior of a classical phantom field usually has an oscillatory character, for which there exist points $\Phi(t_k)=0$, --  the perturbations of the scalar field in which according to \eqref{varphi_sol} grow indefinitely. Figures 1 - 2 show the behavior of the background classical field and its perturbation in such a typical situation.

As is seen from Fig. 2, perturbations of the classical scalar field grow indefinitely at least 3 times in the course of evolution. The perturbations of the energy density of fermions also grow along with them. From a physical point of view, this means instability in a system of scalar charged fermions.

\Fig{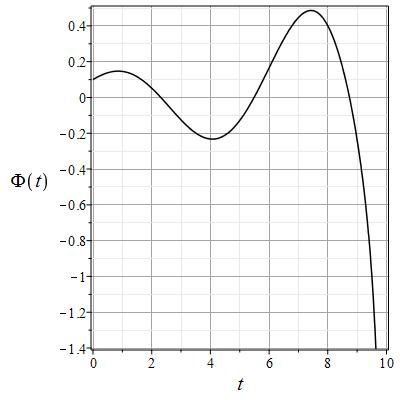}{6}{The evolution of the classical scalar background field $\Phi(t)$ at $\alpha=0.1$, $\beta=0.1$, $m=1$, $\Lambda=0.01$. In this case, there are 3 time instants when it is $\Phi(t_k)=0$.}
\Fig{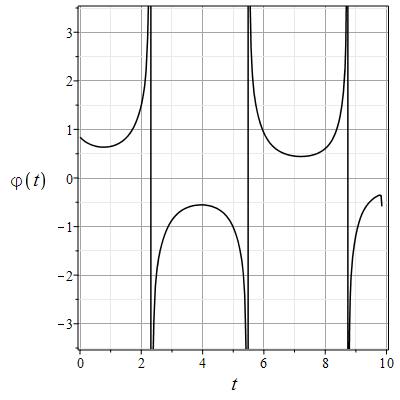}{6}{The evolution of the perturbation amplitude of the classical scalar background field $\Phi(t)$ for the parameter values $\alpha=0.1$, $\beta=0.1$, $m=1$, $\Lambda=0.01$.}

In the case of a phantom field (Fig. 3 - 4), the perturbations of the scalar field, on the contrary, decay over time, due to which the solution asymptotically tends to the background one. Thus, in the case of a phantom field, the system of scalar charged fermions is stable.

The discovered instability of a homogeneous isotropic scalar charged statistical system of degenerate fermions can serve as a mechanism for the formation of a structure in the early Universe, and the formed condensations of cold scalar charged fermion matter can be interpreted as a primary condensate of dark matter.

\Fig{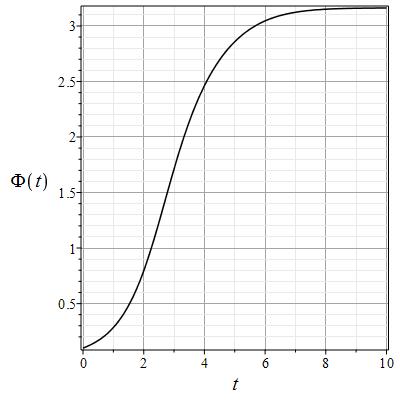}{6}{The evolution of the phantom scalar background field $\Phi(t)$ for the parameter values $\alpha=0.1$, $\beta=0.1$, $m=1$, $\Lambda=0.01$.}
\Fig{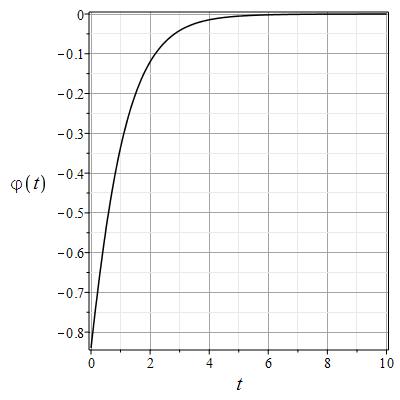}{6}{The evolution of the perturbation amplitude of a phantom scalar background field $\Phi(t)$ for the parameter values $\alpha=0.1$, $\beta=0.1$, $m=1$, $\Lambda=0.01$.}

Let us note, however, that in the neighbourhood of the discovered instability, the condition \eqref{Phi_not_0}, is violated, which leads to the necessity of a more detailed study with the removed strict WKB condition \eqref{extrem_WKB}.

\subsection*{Funding}

 This work was funded by the subsidy allocated to Kazan Federal University for the
 state assignment in the sphere of scientific activities.

\end{document}